\documentclass[letterpaper,12pt]{article}   
\usepackage{osajnl2} 
\usepackage[draft]{hyperref} 
\usepackage{amsmath}
\usepackage{graphicx}
\usepackage{epstopdf}

\begin{document}

\title{Error propagation: a comparison of Shack-Hartmann and curvature sensors }

\author{A. N. Kellerer$^{1,*}$ and A.M. Kellerer$^2$}
\address{$^1$  Institute for Astronomy, \\ 640 N. A'ohoku Place, HI,
Hilo 96720, USA\\
$^*$ Current address: Big Bear Solar Observatory,\\ 
40386 North Shore Lane
Big Bear City, CA, 92314-9672, USA\\ 
kellerer@bbso.njit.edu \\
$^2$ Ludwig-Maximilians University, Munich, Germany}

\begin{abstract} 
Phase estimates in adaptive-optics systems are computed by use of wavefront sensors such as Shack-Hartmann or curvature sensors. In either case the standard error of the phase estimates is proportional to the standard error of the measurements; but the error-propagation factors are different.  We calculate the ratio of these factors for curvature and Shack-Hartmann sensors in dependence on the number of sensors, $n$, on a circular aperture. If the sensor spacing is kept constant and the pupil is enlarged, the ratio increases as $n^{0.4}$. When more sensing elements are accommodated on the same aperture, it increases even faster, viz. proportional to $n^{0.8}$. With large numbers of sensing elements this increase  can limit the applicability of curvature sensors.
 \end{abstract}


\maketitle 

\section{Introduction}
In adaptive-optics (AO) systems  the wavefront sensing is commonly performed with Shack-Hartmann (SH) sensors\,\cite{SH} or Roddier curvature sensors (RC)\,\cite{Roddier1988}. 
Both types of wavefront sensors have been successfully employed in fields such as astronomy\,\cite{Lena} or ophthalmology\,\cite{Liang1997},\cite{Diaz2006}, \cite{Torti2008} -- with up to 188 sensing elements in the case of RC sensors\,\cite{AO188} and nearly a thousand elements in the case of SH sensors\,\cite{AEOS}. 
For the next generation of adaptive optic systems, with increasingly larger numbers of elements, it is of interest to compare -- for different numbers, $n$, of sensing elements -- the two techniques in terms of the error-propagation factor, i.e. the ratio of the standard error of the inferred phase estimates to that of the measured signal.

The error-propagation factor, $e_{\rm SH\/}$, for SH sensors has been found to increase logarithmically with the number, $n$, of sensing elements when their spacing is kept constant and the pupil surface is increased\,\cite{Hudgin}, \cite{Fried}. 
Simulations by N. Roddier have shown that under the same condition, the error propagation, $e_{\rm RC\/}$, for RC sensors increases linearly with $n$\,\cite{N.Roddier}. For a constant pupil surface -- i.e. if more sensing elements are packed on the same aperture --  $e_{\rm RC\/}$ is almost independent of $n$\,\cite{Kellerer}.

Here, the ratio, $e^{RC}/e^{SH}$, of the error-propagation factors for RC and SH sensors is examined in relation to the number of sensing elements, on a circular aperture. The ratio is found to increase as $n^{0.4}$ when the spacing of the sensing elements is fixed. It increases even faster, roughly as $n^{0.8}$, when more sensing elements are placed on a given aperture.

\section{General considerations}\label{general}

In practice the wavefront reconstruction is performed in terms of matrices that are pre-calculated for the particular AO system. A {\it command matrix\/} of voltages is sent to the set of mirror actuators so that in combination with their {\it influence function\/}, i.e. the matrix of  phase shifts caused by a unit shift (voltage increment) of an actuator, the correct {\it reconstruction matrix\/} of phase increments is obtained. In other words, the {\it reconstruction matrix\/} that transforms the measured {\it slope\/} or {\it curvature\/} values into the phase estimates is realized as the `product' of the {\it command matrix\/} and the {\it influence function\/}.
 
While the actual reconstruction algorithms differ, they all contain -- regardless whether they are used with SH- or RC-sensors -- the solution of the Laplace equation in terms of {\it mean-curvature\/} values. Separation of the error propagation into the two inherent steps {\it sensor signal\/} to {\it curvature\/}  and then {\it curvature\/} to {\it phase estimates\/} will, thus, make the analysis more transparent without changing the results. 
 
For an orthogonal grid of sensors the linkage between {\it slope\/} and {\it curvature\/} values is straight-forward. For the common radially symmetric arrangements of curvature sensors it is more complicated. The reconstruction formalism is then still based on the solution of the Laplace equation in terms of the  {\it mean curvatures\/}, and the algorithm that is described in section 2.C (and is employed in sections 4 and 5) will still apply in just slightly modified form. However the separation of the two steps of error propagation is most transparent for an orthogonal pattern of sensors, and this special case will bring out the essential characteristics of error propagation for SH and for RC sensors.

\subsection{Phases,  slopes, curvatures:}

Let the {\it phases\/} at the sensor position $(i,k)$ be $F_{i,k}$. Only the phase {\it differences\/} are of interest.  They are measured in $x$- and $y$-direction by SH sensors and are termed {\it slopes\/}. The relation between the true, i.e. error-free values are:
\begin{eqnarray}
x_{i,k}  = F_{i +1, k}  - F_{i,k}   \nonumber \\	
y_{i,k}  = F_{i, k+1}  - F_{i,k}			\label{eq:1}	            
\end{eqnarray}

The {\it mean curvatures\/} are measured by RC sensors: 
\begin{eqnarray}
c_{i,k}  =  \,(x_{i,k} - x_{i-1,k} + y_{i,k} - y_{i,k-1})  / 4  =  \,(F_{i-1, k} + F_{i+1, k} + F_{i, k-1} + F_{i, k+1})  / 4 - F_{i,k}            \label{eq:2}
\end{eqnarray}
Analogous relations that include less than 4 adjacent sensors apply to the sensors at the boundary of the aperture.

In actuality, the variables $x_{i,k}$ or $y_{i,k}$ are slopes times grid spacing, $d$, and the $c_{i,k}$ are mean curvatures times $d^2$, i.e. all variables are phase increments. To distinguish the variables {\it slope\/} and {\it curvature\/} from the slope and curvature proper they are written in {\it italics\/}, the latter are written in roman. 

\subsection{Iterative reconstruction}\label{linearity}
As shown in earlier analyses (e.g. Hudgin\,\cite{Hudgin}), each phase estimate, $F_{i,k}$, is a linear combination of the elements $x_{l,m}$ and $y_{l,m}$ of the {\it slope\/} matrices:
\begin{eqnarray}
 F_{i,k}  = \sum_{l,m}  \,^x f_{i,k}(l,m)  \,x_{l,m} +   \sum_{l,m} \,^y f_{i,k}(l,m)  \,y_{l,m} 	\label{eq:3}		            
\end{eqnarray}
$^x f_{i,k}$ and $^y f_{i,k}$ are the reconstruction matrices from {\it slopes\/} to phase estimates, $F_{i,k}$. 
With regard to the {\it mean curvatures\/} one can use the same concept, but with different reconstruction matrices : 
\begin{eqnarray}
F_{i,k}  = \sum_{l,m}  f_{i,k}(l,m)  \,c_{l,m}				\label{eq:4}
\end{eqnarray}
Even where {\it slopes\/} are measured they are converted into {\it curvatures\/}, because the less complicated Eq.\,\ref{eq:4} provides the same phase estimates as Eq.\,\ref{eq:3}. The reconstruction matrices $f_{i,k}$ depend on the sensor grid and position; they are pre-calculated for a given AO setup.

A given set of {\it curvatures\/} determines the best phase estimates, i.e. the estimates that are the least-squares fit to the measured {\it curvatures\/} or the underlying {\it slopes\/}. Starting with an arbitrary set, $F_{i,k}$, of estimates the value on every grid point is replaced in each iteration step by the average of the 4 adjacent values, with adjustment for the measured {\it slope\/} or the corresponding {\it curvature\/} data\,\cite{Hudgin}:
\begin{eqnarray}
F_{i,k}  \leftarrow    &(&F_{i-1, k} + F_{i+1, k} + F_{i, k-1} + F_{i, k+1} + x_{i,k} - x_{i-1,k} + y_{i,k} - y_{i,k-1})  / 4 \nonumber \\
  &= &(F_{i-1, k} + F_{i+1, k} + F_{i, k-1} + F_{i, k+1})  / 4 - c_{i,k}		\label{eq:5}
\end{eqnarray}

The new value $F_{i,k}$ fits best the preceding phase estimates at the adjacent 4 grid points and the measured  {\it slope\/} or  {\it curvature\/} data. The analogous relations for sensor positions on the edge of the pupil which have only 1, 2, or 3 neighbors are evident and are not listed. The interior {\it curvature\/} values contain no information on tip and tilt. 
The procedure converges to the least-squares solution. Depending on $n$, several hundred iterations may be required, but the procedure converges correctly. 

\subsection{Reconstruction by matrix inversion}\label{sec:2c}
The iterative procedure is simple and transparent, but the direct solution by inversion of one large matrix that connects phases and {\it mean curvatures\/} is a convenient alternative. To employ it  one arranges the variables for the different sensor positions as vectors, rather than matrices. This is merely a change of notation; instead of dealing with a matrix of matrices as in Eq.\,\ref{eq:4} one deals with a single matrix equation:
\begin{eqnarray}
c = U\cdot F 
\end{eqnarray}

In this alternative description $F$ and $c$ are the vectors of the phase values and the {\it mean curvatures\/}. They contain $n$ elements, where $n$ is the number of sensors. $x$ and $y$ is the analogous notation for the {\it slopes\/}. The $n \times n$  matrix  $U$ contains in each row at the appropriate positions the  term -1 and four terms 1/4 when the phase corresponds to an interior position. For positions on the boundary of the sensor grid there are, apart from -1, three terms 1/3, two terms 1/2, or one term 1.

If all phases are equal, the {\it curvature\/} vector, $c$, is zero, which shows that the matrix $U$ is singular. The reconstruction relation contains, therefore, instead of the familiar inverse the pseudo-inverse, $R$, of $U$:

\begin{eqnarray}
F = R\cdot c  \label{eq:FRc} 
\end{eqnarray}

Eq.\,\ref{eq:FRc} is the equivalent of Eq.\,\ref{eq:4}. The column of the matrix $R$ that corresponds to the sensor position $(l,m)$ in Eq.\,\ref{eq:4} is a vectorial representation of the matrix  $f_{i,k}(l,m)$.  Also it is readily seen that the equivalent of Eq.\,\ref{eq:3} is :

\begin{eqnarray}
F = \,^x R\cdot x  + \,^y R\cdot y \label{eq:FRxy} 
\end{eqnarray}

where each column in the two matrices  $^xR$   and  $^yR$  is the difference of  the two columns in $R$ that correspond to the pair of sensor positions associated with the {\it slope\/} $x$ or $y$.

The vectorial representation of the {\it slopes, mean curvatures\/} and phases is utilized only in section 5, where Eqs.\,\ref{eq:FRc} and \ref{eq:FRxy} are used to obtain, for sensor numbers up to $n=500$, the error-propagation factors for RC and SH sensors. The use there of the inversion algorithm is a matter of convenience. In the absence of tools to handle very large matrices the iterative method serves the same purpose.

\subsection{Error propagation}
The different types of  wavefront sensors have particular strengths and limitations that determine their applicability. The present discussion is directed at one characteristic of SH and RC sensors: the different propagation of measurement errors to phase-estimation errors. 

Let $z$ and $v$ be the measured signals of the SH and RC sensors; $z$ and $v$ are dimensionless values that can -- with coefficients specified in the subsequent section -- be converted to the {\it slope\/} and {\it mean curvature\/} values. 

If $\sigma_z$ and $\sigma_v$ are the standard errors of the SH and RC signals, $z$ and $v$, then the phase errors, $\sigma_{F, SH}$ and $\sigma_{F, RC}$, are the products of the errors $\sigma_z$ and $\sigma_v$ and the {\it error-propagation factors\/}, $e_{SH}$ and $e_{RC}$: 

\begin{eqnarray}
\sigma_{F, SH} = \sigma_z\cdot e_{SH}  \nonumber \\
\sigma_{F, RC} = \sigma_v\cdot e_{RC}  \label{eq:sigmaf}
\end{eqnarray}

\begin{figure}[htbp]
\begin{center}
\includegraphics[width=1\textwidth]{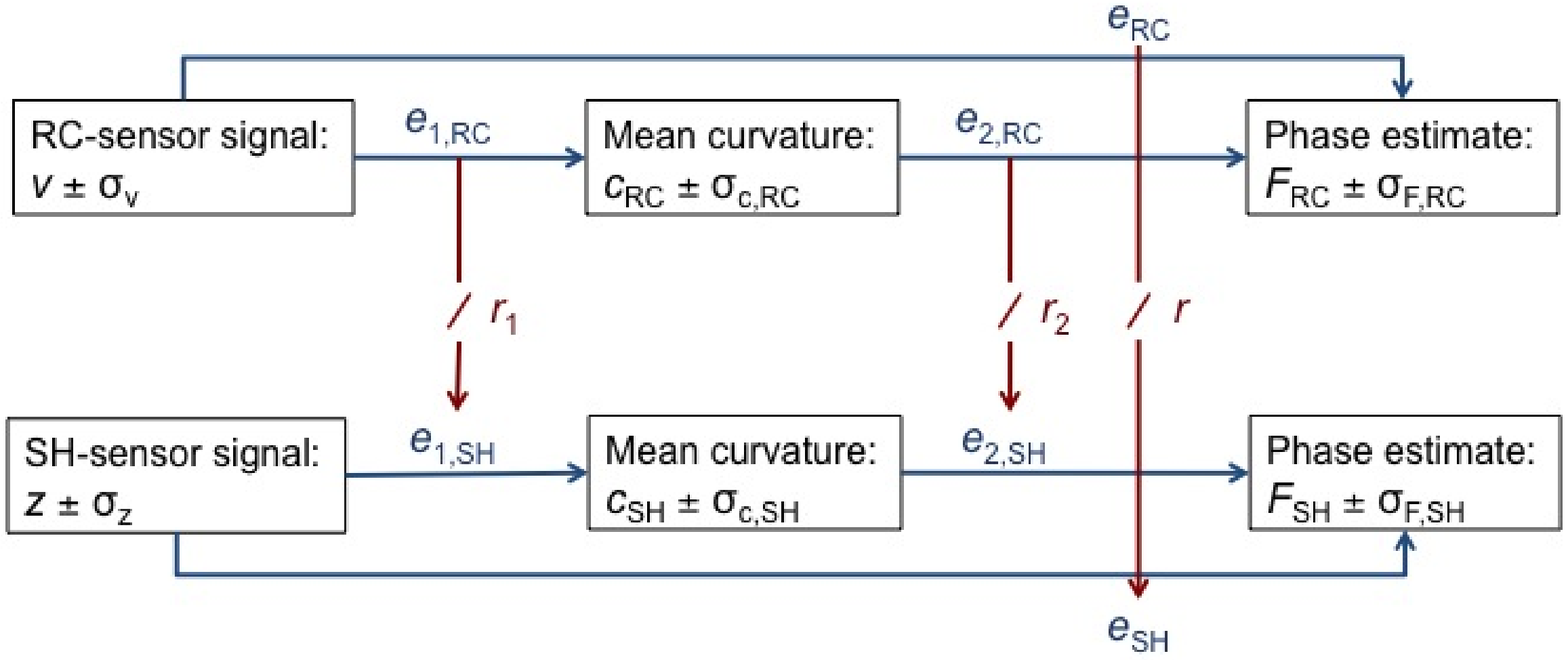}
\caption{The analysis of the error propagation in SH- and RC-sensors is separated into two steps:  signals to mean curvatures and mean curvatures to phase estimates.}
\label{fig:scheme}
\end{center}
\end{figure}

The exploration of these error-propagation factors will, as indicated in Fig.\,\ref{fig:scheme}, be separated into two steps. The first step considers the two fairly trivial partial factors from $\sigma_z$ and $\sigma_v$ to the standard errors of the {\it mean curvatures\/}:
\begin{eqnarray}
e_{1,SH} = \sigma_{c, SH} / \sigma_z \nonumber \\
e_{1,RC} = \sigma_{c, RC} / \sigma_v  
\end{eqnarray}

This step requires the specification of the calibration factors from SH- or RC-sensor signal to {\it mean curvature\/}, which will be detailed in the next section. 

The second step -- which is the major part of this study -- is the determination of the propagation factors from the {\it mean curvature\/} errors to the phase errors, $\sigma_{F, SH}$ and $\sigma_{F, RC}$:
\begin{eqnarray}
e_{2,SH} = \sigma_{F, SH} / \sigma_{c, SH}  \nonumber \\
e_{2,RC} = \sigma_{F, RC} /  \sigma_{c, RC}   
\end{eqnarray}

The error propagation for RC and for SH sensors will be compared in terms of the ratio:
\begin{eqnarray}
r = e_{RC} / e_{SH} = e_{1,RC} / e_{1,SH} \cdot e_{2,RC} / e_{2,SH} = r_1\cdot r_2
\end{eqnarray}

 $e_{1,RC}, e_{1,SH}$ and $r_1$ are considered in the next section in relation to the number, $n$, of sensors. Sections \ref{sec:analytic} and \ref{sec:sim} will subsequently deal with the second step, the error propagation from {\it mean curvature\/} to phase.

\section{Comparison of the curvature errors from SH and RC sensors}\label{sec:r1}

\subsection{SH sensors}

The signal, $z$, equals the measured displacement of the image in the detector plane -- in units of pixel size, $p$.  
The relation between the {\it slope\/} value $s$, i.e. $x_{i,k}$ or $y_{i,k}$, and the signal, $z$, of the SH sensor is\,\cite{SH}: 
\begin{eqnarray} 
s=\frac{2\pi}{\lambda}\,\frac{d\,p}{f_l}\,z
\end{eqnarray}

where $\lambda$ is the wavelength, $f_l$ the focal distance of the SH lenslets, and $d$ the grid spacing. 

The same relation applies to the standard errors, $\sigma_s$ and $\sigma_z$, of $s$ and $z$. To obtain the expression for $e_{1,SH}$ one needs, in addition, the relation between $\sigma_s$ and the standard error, $\sigma_{c, SH}$, of the {\it mean curvature\/} derived from the {\it slope\/} values. This relation is simple, since the standard error of the  {\it mean curvature\/} equals -- for internal sensors -- half the standard error of the {\it slope\/} values (see Eq.\,\ref{eq:2}). For the sensors on the boundary the value is somewhat larger, but averaged over all sensors the difference can be disregarded -- especially for large $n$. With the factor 1/2 one obtains: 

\begin{eqnarray}
e_{1,SH} = \sigma_{c,SH} /\sigma_s = \pi\,d\,p / \lambda f_l
\end{eqnarray}

when $n$ is increased, while the grid distance, $d$, and the size of the sub-apertures remain the same, $\lambda,\,p$ and $f_l$ are kept constant. The error-propagation factor, $e_{1,SH}$, is then independent of $n$.

On the other hand, when more sensors are packed on the same aperture, the focal length, $f_l$, of the Shack-Hartmann lenses is generally increased in order to keep the amplitude of the image motions constant in the detector plane. Since the amplitude of the sensed phase distortions decreases with sub-aperture size as $d^{5/3}\sim n^{-5/6}$ \cite{Roddier1981}, $f_l$ is adjusted proportional to $n^{5/12}$. One concludes that the error propagation from the SH sensor signal to {\it mean curvature\/} is then roughly inverse to the number of sensors:
\begin{eqnarray}\label{e1sh}
e_{1,SH}  \propto n^{-11/12} 
\end{eqnarray}

\subsection{Curvature sensors}

The relation between the {\it mean curvature\/}, $c$, and the signal, $v$, of the RC sensors is\,\cite{Roddier1988}: 
\begin{eqnarray}
c=\frac{2\pi}{\lambda}\,\frac{d^2\,l}{f^2}\,v
\end{eqnarray}
Accordingly:
\begin{eqnarray}
e_{1,RC} =\sigma_{c,RC}/\sigma_{v} = \frac{2\pi}{\lambda}\,\frac{d^2\,l}{f^2}
\end{eqnarray}

Again, the error-propagation factor is independent of $n$, when $n$ is increased, while the grid distance, $d$, and the size of the sub-apertures remain the same. 

When more sensors are packed on the same aperture, the telescope focal length, $f$, remains unchanged, but the extra-focal distance, $l$, of the curvature sensor is increased proportional to $n^{1/2}$, so that the size $d\,l/f$, of the sensing elements in the detector plane stays constant. 
It follows, that
\begin{eqnarray}\label{e1rc}
e_{1,RC}  \sim  n ^{-1/2} 
\end{eqnarray}

Eqs.\,\ref{e1sh} and \ref{e1rc} imply that the ratio, $r_1$, increases when more sensors, $n$, are packed on the same aperture:
\begin{eqnarray}
r_1 = e_{1,RC}/e_{1,SH} \sim n^{5/12}
\end{eqnarray}

\section{Error propagation from curvature to phase -- Comparison of SH and RC sensors}\label{sec:analytic}

It is not surprising that the error propagation from {\it curvature\/} to phase estimation should be somewhat worse for the directly determined {\it mean curvatures\/}. 
Broadly speaking, a measurement error of the {\it slope\/} at a sensor position causes an imprecise estimate of the local phase increment only, while an error of the measured {\it curvature\/} causes an imprecise estimate of the rate of phase increments, which then affects also distant estimates of phase increments. This is analogous to reconstructing a function of a single variable from its imprecisely measured second derivative, which is less stable than reconstructing it from the imprecisely measured first  derivative. 
While there is, thus, no particular need to explain the difference, it is still helpful to consider the issue in analytical terms before numerical results are derived in section\,\ref{sec:sim}.

\subsection{Correlated vs. uncorrelated curvatures }

The observed {\it slopes\/} or {\it mean curvatures\/} and the phase estimates are the sum of the true values and the errors. Since the reconstruction is additive, the errors of the phase estimates depend -- regardless of the true {\it slope\/}, {\it curvature\/} and phase values -- only on the measurement errors. The subsequent considerations are therefore simplified by taking the true values to be zero, so that the {\it slope\/} and {\it mean curvature\/} values  $x_{l,m}$, $y_{l,m}$ and $c_{l,m}$ stand for the errors only. 

To facilitate the comparison of error propagation, two different procedures will be  invoked that generate sets of {\it mean curvatures\/} as obtained with SH and RC sensors. Both procedures start out by simulating  independent and unbiased random {\it slope\/} errors; the mean square of these errors is set equal to 4, so that the mean square of the corresponding $c_{l,m}$ is unity. The boundary effect is here disregarded, i.e. all measured {\it curvatures\/} are treated as interior values that are linked to 4 {\it slope\/} values. The boundary could be accounted for, but this aspect is here not essential.  
 
Procedure A corresponds to the use of SH sensors; it generates one set of $2n$ {\it slope\/} errors, $x_{l,m}$, $y_{l,m}$, and derives from these values, in line with Eq.\,\ref{eq:2}, the set of {\it mean curvatures\/}, $c_{l,m}$.  Adjacent values $c_{l,m}$ are then to some degree correlated.
 
Procedure B simulates {\it separately\/} for each  position, $(l,m)$, the four associated random errors $x_{l -1,m}, y_{l-1,m}, x_{l,m}$, and $y_{l,m}$, and derives from them the {\it curvature\/} error,  $c_{l,m}$ . As with procedure A, the resulting mean square {\it curvature\/} error, $<c^2>$, is  unity, but the $c_{l,m}$ are now uncorrelated, i.e. they resemble a set of {\it mean curvatures\/} obtained with RC sensors.
 
Consider first the case of merely one single {\it slope\/} error, say $x_{n,m} = \epsilon$. With procedure A  this error,  $x_{n,m} = \epsilon$,  corresponds to the two {\it mean curvature\/} errors, $c_{n,m} = \epsilon/4$ and $c_{n+1,m} = - \epsilon/4$.  The addition of these two terms according to  Eq.\,\ref{eq:4} gives the associated phase errors:
\begin{eqnarray}
F_{i,k}   =  \epsilon/4	\cdot (f_{i,k}(l,m)  -  f_{i,k}(l+1,m) ) 
\end{eqnarray}

One obtains the corresponding mean square phase error:
\begin{eqnarray}
\partial_{SH}^2   &=&  \epsilon ^2/ 16 \cdot  [ < (f (l,m) - f(l+1,m))^2 > - < (f (l,m) - f(l+1,m)) >^2 ] \nonumber \\
	          &=&   \epsilon ^2/ 16 \cdot [ < f(l,m)^2 > - < f(l,m) >^2  + <  f(l+1,m)^2> - <  f(l+1,m)>^2  \nonumber  \\
	           &&     - 2 (< f(l,m) \cdot f(l+1,m) > - < f(l,m) > \cdot < f(l+1,m) > ) ] 
\end{eqnarray}
$<..>$ denotes the average over all $(i,k)$. 

With the notation $var(\, )$ and $cov (\,,\,)$ for $variance$ and $covariance$ and with $<\epsilon^2>=1$, this takes the form:
\begin{eqnarray}
\partial_{SH}^2  =   [var( f(l,m) )   + var( f(l+1,m) ) - 2 cov (f(l,m), f(l+1,m)) ]  / 4	\label{eq:dsh}
\end{eqnarray}          

The total mean square phase error, $\sigma^2_{F,SH}$, is obtained by summing Eq.\,\ref{eq:dsh} for all $2n$ {\it slopes\/}. The contributions of the {\it slopes\/} $x_{l,m}$ and $y_{l,m}$  being equal, it is sufficient to sum Eq.\,\ref{eq:dsh} over all $n$ positions $(l,m)$ and then multiply the result by 2 :  
\begin{eqnarray}
\sigma_{F,SH}^2  = e_{2,SH}^2 = \sum_{l,m}  [ var( f(l,m) )  - cov (f(l,m), f(l+1,m) ]  \label{eq:e2sh2}
\end{eqnarray}          

$\sigma_{F,SH}$  equals the squared error-propagation factor $e_{2,SH}$, because $\sigma_{c,SH} =1$.

With procedure B the single {\it slope\/} error $x_{n,m}$  corresponds to two {\it curvature\/} errors, $c_{l,m} = \epsilon_1/4$   and  $c_{l+1,m} = - \epsilon_2/4$.  Both for $\epsilon_1$ and $\epsilon_2$ the mean square error is $<\epsilon^2>$, but the mixed term $< \epsilon_1\cdot \epsilon_1 >$ in the equation below  vanishes because of the  statistical independence  of $\epsilon_1$ and $\epsilon_2$.  The increment of the  mean square phase error is thus:
\begin{eqnarray}
\partial_{RC}^2  &=&  \epsilon_2^2/16 \cdot [ < f(l,m)^2 > - < f(l,m) >^2 ]   + \epsilon_1^2 /16\cdot[<  f(l+1,m)^2> - <  f(l+1,m) >^2 ]\nonumber \\
	   && - \epsilon_1\, \epsilon_2 / 8 \cdot [< f(l,m) \cdot f(l+1,m) > - < f(l,m) > \cdot < f(l+1,m) > ] 
\end{eqnarray}   

Since the mixed terms disappear one obtains instead of Eq.\,\ref{eq:e2sh2}:
\begin{eqnarray}
\sigma_{F,RC}^2 =   e_{2,RC}^2 =   \sum_{l,m}  var( f(l,m) )   	\label{eq:e2rc2}
\end{eqnarray}   

This is a notable result: The error-propagation factors differ by the covariance term in Eq.\,\ref{eq:e2sh2}.  This term converges with increasing $n$ towards the variance, which keeps down the error propagation, when the number of SH sensing elements becomes larger. In  the case of independent {\it curvatures\/} the covariance term is absent. This increases the error propagation with RC detectors and makes it roughly proportional to $n$, i.e. to the number of terms in Eq.\,\ref{eq:e2rc2}. 

\subsection{Analytical approximation}\label{sec:4B}
The above conclusion,  the poorer error propagation for {\it curvature\/} measurements with large sensor numbers,  will now be quantified. For this purpose a large grid of sensors is considered that extends beyond a central region of $n$ sensor positions for which the phase estimates are actually sought. This {\it large-grid approximation\/} is informative because it entails a simple reconstruction function, $f_{i,k}$, that -- regardless of the sensor position, $(l,m)$ -- is roughly  proportional to the logarithm of the distance between sensor positions.

While the {\it large-grid approximation\/}  brings out the characteristic difference between the use of {\it curvatures\/} from SH and RC detectors, it provides results that differ somewhat from the more accurate numerical evaluation in section\,\ref{sec:sim}.

With grid spacing $d$ the distance of position $(i, k)$ from the central position, $(0,0)$, is:
\begin{eqnarray}
d\,\sqrt{i^2+k^2} = d\cdot u  
 \end{eqnarray}
 
The function $f(x,y)=a\,\ln(u)+b$ has mean curvature zero for $u>0$, i.e. its Laplacian -- or twice the mean curvature -- vanishes: $
\Delta f = \frac{1}{u}\,\frac{\partial}{\partial u} (u\,\frac{\partial f}{\partial u}) =0$.
 
$f(x,y)$ is, therefore -- at least at somewhat larger $u$, where the difference quotient equals the differential quotient -- a solution for the reconstruction function. The constant term, $b$,  is irrelevant and can be disregarded. This is in line with the observation that the reconstruction function is independent of $d$, which means that for given  {\it curvature\/} errors the phase errors are the same, regardless whether the grid spacing, $d$, is decreased in order to accommodate a larger number of sensors, or whether $d$ is kept constant and the aperture is enlarged.

Numerical evaluation in terms of the iterative method shows that:
\begin{eqnarray}
f_{i,k} = 2/\pi\ \ln(u) = \ln(i^2+k^2)/\pi \hskip .3cm {\rm and \/} \hskip .3cm f_{0,0}=f(0)=-1   \label{eq:fik}
\end{eqnarray}
 applies exactly at $u=0, u=1$ and for large $u$. For intermediate $u(>1)$ the difference to the exact solution is too small to be of concern for the present approximation. 
 
Eq.\,\ref{eq:fik} gives the reconstruction function, $f_{i,k}(0,0)$, for the central sensor position. For position $(l,m)$ it is shifted accordingly:
 \begin{eqnarray}
 f_{i,k} (l,m) = \ln((i-l)^2+(k-m)^2)/\pi \hskip .3cm {\rm and \/} \hskip .3cm f_{l,m}(l,m)=-1   \label{eq:fiklm}
 \end{eqnarray}
 
Fig.\,\ref{fig:1} gives the error-propagation factors $e_{2,RC}$ and $e_{2,SH}$ that are obtained by inserting the above expression into Eqs.\,\ref{eq:e2sh2} and \ref{eq:e2rc2}, where the variance and covariance are -- again with the symbol $<\,>$ for average over all $(i,k)$:  
 \begin{eqnarray}
 var( f(l,m)) &=& < f_{i,k} (l,m)^2> - <f_{i,k}(l,m)>^2 \nonumber \\
 cov(f(l,m),f(l+1,m)) &=& < f_{i,k} (l,m)\cdot f_{i,k} (l+1,m)> \nonumber \\  && - < f_{i,k} (l,m)>\cdot < f_{i,k} (l+1,m)>
  \end{eqnarray}

The results are given for symmetric circular sets of sensors on an orthogonal grid. The center of the grid  is on the midpoint between 4 sensors ( $n = 12, 16, 24, 32, ..$). 
The main finding is the very slow increase with $n$ of the error propagation from {\it curvature\/} to phase in the case of the indirectly measured {\it curvatures\/} (SH), versus the more substantial increase for the directly measured {\it curvatures\/} (RC). 

\begin{figure}[htbp]
\begin{center}
\includegraphics[width=0.5\textwidth]{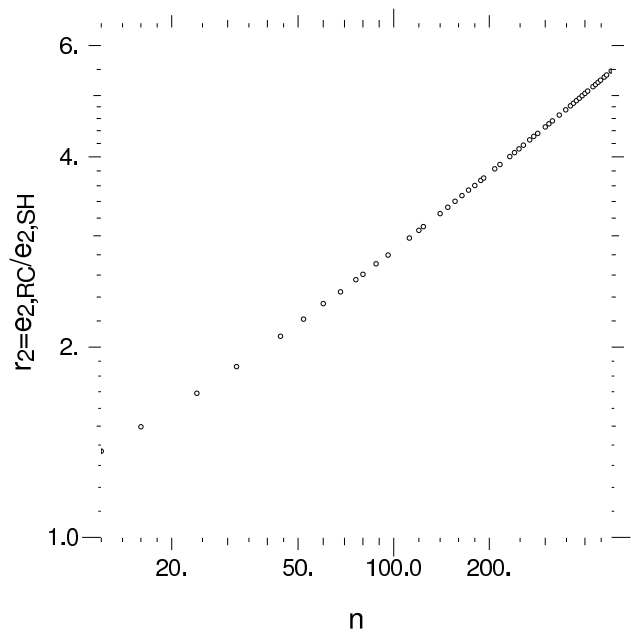}
\caption{Ratio $r_2 = e_{2,RC} / e_{2,SH}$. The  results are given for the number $n$ of sensors that occur within circular regions, with the grid-center positioned in between 4 sensors. }
\label{fig:2}
\end{center}
\end{figure}

Fig.\,\ref{fig:2} gives the corresponding ratio $r_2 = e_{2,RC} / e_{2,SH}$  of the error-propagation factors. For sensor numbers up to about 40, $r_2$  is less than 2, for $n = 500$ it is roughly 5.5.

\section{Explicit computation of the error propagation from curvature to phase}\label{sec:sim}

Eqs.\,\ref{eq:FRc}  and \ref{eq:FRxy}  can, largely in analogy to the preceding section, be used to derive exact values of the error-propagation factors  $e_{2,RC}$ and $e_{2,SH}$. Since the {\it mean curvature\/} errors from RC detectors are statistically independent, their mean-square contributions to the phase errors add up. The contribution of the sensor labeled $k$ in the vector representation equals therefore the product of the mean-square {\it curvature\/} error $<\epsilon_c^2>$ and the  sum of the squared components of column $k$ in the reconstruction matrix, $R$. Since the contributions of all sensors add up,  the mean squared phase error equals the sum of all squared components of $R$. The error-propagation factor is thus:

\begin{eqnarray}
e_{2,RC} = n^{0.5}\, < R_{i,k}^2 >    \label{eq:31}
\end{eqnarray}

To compute the error propagation with SH detectors one must go back -- as in the considerations in the previous section -- to the {\it slope\/} errors, i.e. one needs to use Eq.\,\ref{eq:FRxy} instead of Eq.\,\ref{eq:FRc}. Each of the  columns of the matrices   $^xR$  and $^yR$ are, as pointed out in section\,\ref{sec:2c}, the differences of two associated columns of $R$. The sum of the squared matrix elements is, therefore, in analogy to Eq.\,\ref{eq:e2sh2}, the sum of the squares of the elements of $R$ minus a sum of the product of associated elements. This provides the relation:
\begin{eqnarray}
e_{2,SH} &=& n^{0.5}\, ( <  \,^xR_{i,k}^2 > + <  \,^xR_{i,k}^2 > ) / 4  \nonumber \\
&=&  n^{0.5} \, < \,^xR_{i,k}^2 > / 2   	\label{eq:32}
\end{eqnarray}

\begin{figure}[htbp]
\begin{center}
\includegraphics[width=0.5\textwidth]{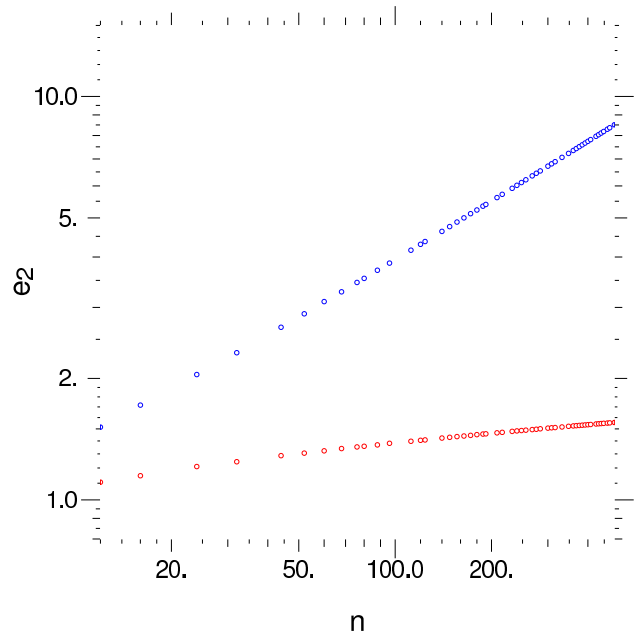}
\caption{Upper line of symbols: $e_{2,RC}$. Lower line of symbols: $e_{2,SH}$.  The  results are given for the number $n$ of sensors that occur within circular regions, with the grid-center positioned in between 4 sensors.}
\label{fig:1}
\end{center}
\end{figure}

\begin{figure}[htbp]
\begin{center}
\includegraphics[width=0.5\textwidth]{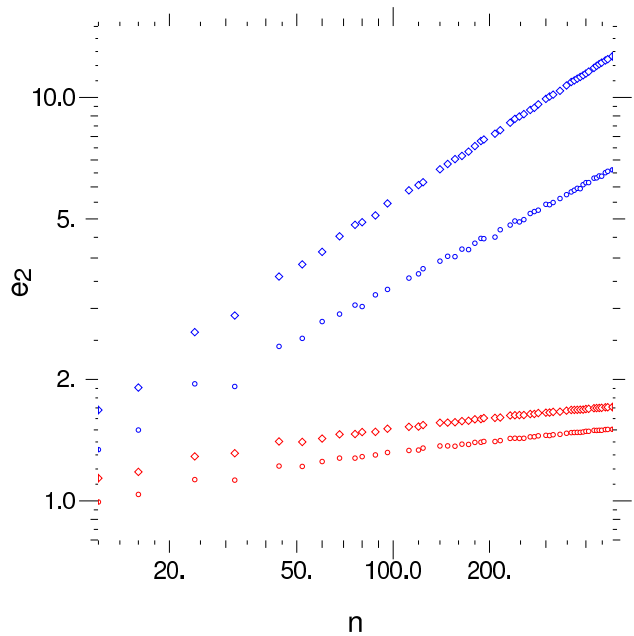}
\caption{Error propagation factors $e_{2,RC}$ (upper pair of curves) and $e_{2,SH}$  (lower pair of curves). In each pair of curves the upper curve gives the total $e_2$-factor, while the lower curve gives the reduced value that remains when tip and tilt are removed from the estimates.  The results  without tip and tilt compare to the ones in Fig.\ref{fig:1}. -- The results are obtained for an orthogonal grid of $n$ sensors within a circular aperture. The grid center is positioned in between 4 sensors.}
\label{fig:3}
\end{center}
\end{figure}

The diamonds in Fig.\,\ref{fig:3} give the results. While the dependences on $n$ agree essentially with those in Fig.\,\ref{fig:1}, the absolute values of $e_{2,RC}$ and $e_{2,SH}$ are larger than in Fig.\,\ref{fig:1}.  This is not surprising, because the approximation in section\,\ref{sec:4B} has inherent symmetry that excludes tip and tilt. For a further comparison it is therefore informative to determine the phase errors with tip and tilt correction. 
They are given in Fig.\,\ref{fig:3} by the smaller dots. 
The correction for tip and tilt reduces the phase-estimation errors substantially, especially for  the RC data. 

The contribution of tip and tilt is seen to almost double the standard error of the phase estimates in the case of directly measured {\it curvatures\/}, which means that it contributes  predominantly to the variance of the phase-estimation errors. For {\it curvature\/} values derived in terms of SH sensors the relative contribution of tip and tilt is markedly less; it contributes about half of the variance.

The analytical approximation in section\,\ref{sec:4B} has been solely based on the {\it mean curvatures\/}. Since these carry no information on tip and tilt, the curves in Fig.\,\ref{fig:1} should correspond to the results in Fig.\,\ref{fig:3} without tip and tilt. They run, in fact, below the curves that include tip and tilt, but lie above those without tip and tilt, which reflects inaccuracies due to the large grid-approximation. 

Overall, however, there is reasonable agreement, and the essential finding is, that the curves for the RC detectors in Figs.\,\ref{fig:1} and \ref{fig:3} have the same slope of about 0.5,  which means that $e_{2,RC}$  is proportional to  $n^{0.5}$. The curves for the SH detectors exhibit a slope close to 0.1, i.e. they show that $e_{2,RC}$ increases roughly proportional to $n^{0.1}$; over the relevant range of $n$  the slow increase is equally in accord with a logarithmic dependence as earlier reported by Hudgin\,\cite{Hudgin} and by Fried\,\cite{Fried}.

\begin{figure}[htbp]
\begin{center}
\includegraphics[width=0.5\textwidth]{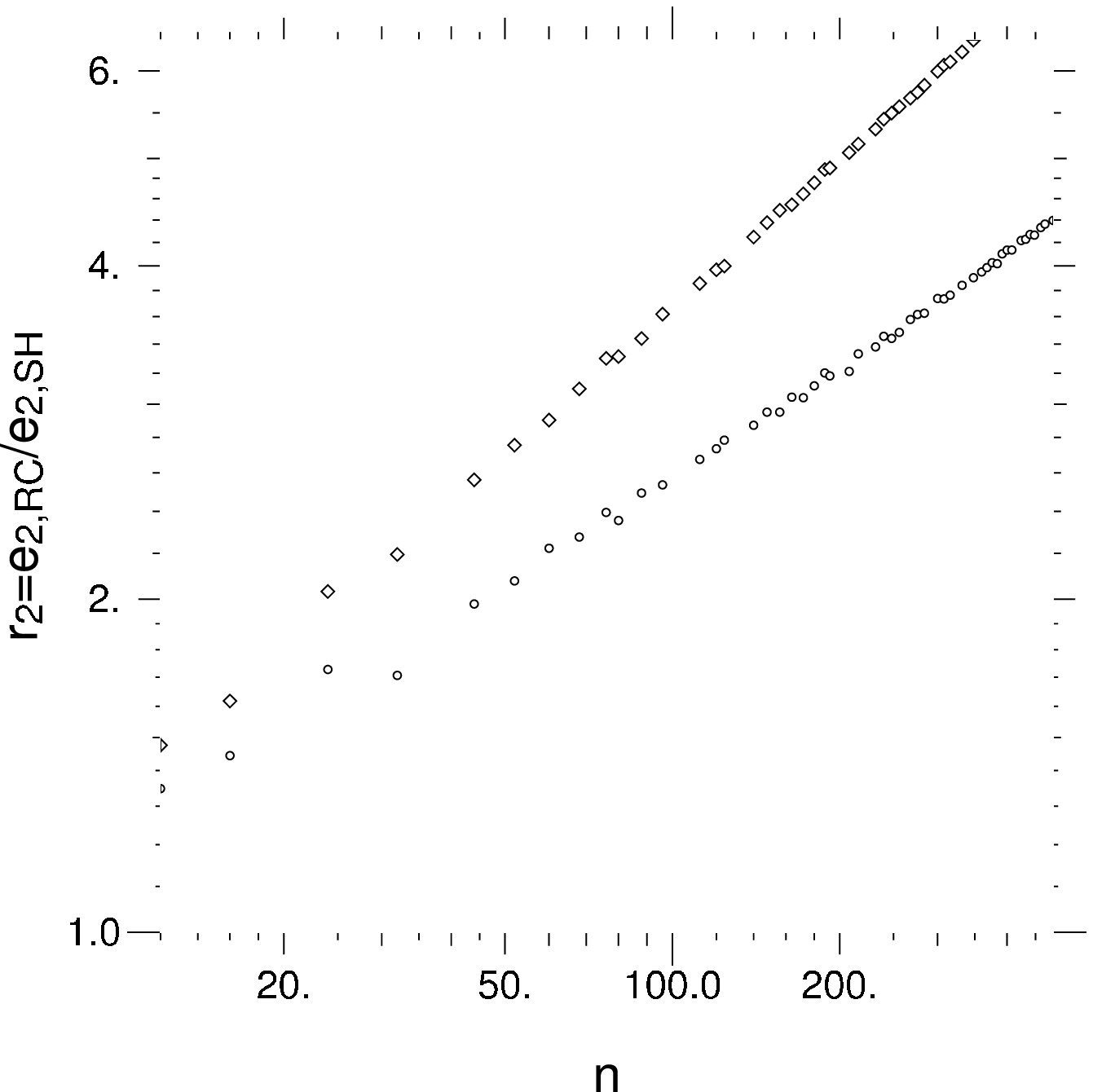}
\caption{The ratio, $r_2$, of the error-propagation factors $e_{2,RC}$ and $e_{2,SH}$ represented  in Fig.\,\ref{fig:3}. The upper line of symbols relates to the entire standard error of the phase estimates. The lower line of symbols relates to the standard error without the contribution of tip and tilt. It is in reasonable agreement with the approximate results represented in Fig.\,\ref{fig:2}. }
\label{fig:4}
\end{center}
\end{figure}

Fig.\,\ref{fig:4} gives the ratios $r_2$ that corresponds to Fig.\,\ref{fig:3}. With tip and tilt excluded  $r_2$ reaches the value 2  at $n =50$ and exceeds 4 at $n=500$.  The values without tip and tilt correction are about 1.5 times larger, but they are not highly relevant, because in AO the use of RC detectors is commonly combined with measurements to allow tip and tilt correction.

\section{Conclusion}
Shack-Hartmann {\it slope\/} detectors and Roddier {\it curvature\/} sensors have particular strengths and limitations that determine their applicability. The present analysis has been concerned with one characteristic of the wavefront sensors, the propagation of measurement errors to phase-estimation errors. With either type of sensor the standard error of the phase estimates is proportional to the standard error of the {\it curvature\/} values, but the error-propagation factors differ. The errors of the directly measured {\it curvatures\/} are statistically independent, and therefore their contributions to the mean-square phase errors add up. In contrast, when the mean {\it curvature\/} values are obtained from {\it slope\/} measurements, adjacent values are negatively correlated and accordingly their  mean-square phase-estimation errors combine less than additively.

The results confirm and quantify the earlier findings that the error propagation is larger when the {\it curvatures\/} are measured directly instead of being derived from measured {\it slopes\/}. The ratio, $e_{RC}/e_{SH}$, of the error-propagation factors for {\it curvature\/} and {\it slope\/} measurements is seen to increase roughly as $\sqrt{n}$, when the size of the sensing elements is fixed and the pupil size is enlarged. When more sensing elements are packed onto  the same pupil, the ratio increases even faster; $e_{RC}/e_{SH}$  is then roughly proportional to $n$. 

Finally it needs to be noted that the standard errors of the phase estimates are the product of the standard errors of the detector response  and the error-propagation factor (see Eq.\,\ref{eq:sigmaf}). For photon-noise limited measurements, the standard error of the detector response is inversely proportional to the square root of the  mean number of detected photons, i.e. the product of fluence and the area of the sub-apertures. Since the photon number is inversely proportional to $n$, one concludes that the standard errors of the phase estimates increase even faster with $n$  than the error-propagation factors. 

The difference in error propagation is not decisive at small or moderate sensor numbers, $n$, and, generally, the error propagation is by no means the only or even the dominant aspect of the applicability of the different types of sensors. The substantial increase of the error-propagation factor for {\it curvature\/} detectors at large $n$ can nevertheless be a critical factor that needs to be weighed against their various advantages. 

\section*{Acknowledgments}
This research was done while the first author worked at the Institute for Astronomy in Hawaii under financial support from the United States Air Force Office of Scientific Research.  Continued guidance by  Mark Chun and Christ Ftaclas  is greatly appreciated.

\clearpage

\begin{thebibliography}{99}

\bibitem{SH} 
B. C. Platt, R. Shack, ``History and principles of Shack-Hartmann wave-front sensing'', 
Journal of Refractive Surgery, vol. 17, p. 573-577 (2001).

\bibitem{Roddier1988} 	
F. Roddier, ``Curvature sensing and compensation: a new concept in adaptive optics'', 		
Applied Optics, vol. 27, issue 7, p.1223-1225 (1988)

\bibitem{Lena} 	
P. Lena, 
``Adaptive optics: a breakthrough in astronomy'',
Experimental Astronomy, vol. 26, p.35-48 (2009)


\bibitem{Liang1997} 
J. Liang, D. R. Williams, D. T. Miller, 
``Supernormal vision and high-resolution retinal imaging through adaptive optics'', 
J. Opt. Soc. Am. A, vol. 14, p. 2884-2892 (1997)

\bibitem{Diaz2006} 
F. Diaz-Douton, J. Pujol, M. Arjona, S. O. Luque,
``Curvature sensor for ocular wavefront measurement'', 
Optics Letters, vol. 31, i. 15, p. 2245-2247 (2006)

\bibitem{Torti2008}
C. Torti, S. Gruppetta, L. Diaz-Santana,
``Wavefront curvature sensing for the human eye''
Journal of Modern Optics, vol. 55, i. 4-5, p. 691-702 (2008)

\bibitem{AO188}
M. Watanabe, S. Oya, Y. Hayano, H. Takami, M. Hattori, Y. Minowa, Y. Saito, M. Ito, N. Murakami, M. Iye, O. Guyon, S. Colley, M. Eldred, T. Golota, M. Dinkins, 
``Implementation of 188-element Curvature-based Wavefront Sensor and Calibration Source Unit for the Subaru LGSAO System'', 
Proc. SPIE, vol. 7015, p. 701564-701564-8 (2008)

\bibitem{AEOS} L. C. Roberts, C.R. Neyman, 
``Characterization of the AEOS Adaptive Optics System'', 
Publications of the Astronomical Society of the Pacific, vol. 113, p. 1260-1266 (2002)

\bibitem{Hudgin}  R. Hudgin,
``Wave-front compensation error due to finite corrector-element size'',
Optical Society of America, Journal, vol. 67, p. 393-395 (1977)

\bibitem{Fried}  D. Fried,
``Least-square fitting a wave-front distortion estimate to an array of phase-difference measurements'', 
Optical Society of America, Journal, vol. 67, p. 370-375 (1977)

\bibitem{N.Roddier} 
N. Roddier, ``Curvature sensing for adaptive optics: a computer simulation'', thesis for the degree of master of science submitted to the University of Arizona (1989)

\bibitem{Kellerer} 
A. Kellerer, ``Curvature sensors: noise and its propagation'', 
JOSA A, vol. 27, issue 11, p.A29-A36 (2010)

\bibitem{Roddier1981} 	
F. Roddier, ``The effects of atmospheric turbulence in optical astronomy'', 		
Progress in optics, vol. 19, Amsterdam, North-Holland Publishing Co., p. 281-376 (1981)
\end{thebibliography}
\end{document}